# An explainable two-dimensional single model deep learning approach for Alzheimer's disease diagnosis and brain atrophy localization

**Fan Zhang** [a, b], **Bo Pan** [d], **Pengfei Shao** [a, b], **Peng Liu** [a, b], **Alzheimer's Disease Neuroimaging Initiative**[1], **the Australian Imaging Biomarkers and Lifestyle flagship study of ageing**[2], **Shuwei Shen** [c, d, *], **Peng Yao** [b, *], **Ronald X. Xu** [a, c, e, *]

[a] Department of Precision Machinery and Precision Instrumentation, University of Science and Technology of China, Hefei 230026, China
[b] Key Laboratory of Precision Scientific Instrumentation of Anhui Higher Education Institutes, University of Science and Technology of China, Hefei 230026, China
[c] Suzhou Advanced Research Institute, University of Science and Technology of China, Suzhou 215000, China
[d] First Affiliated Hospital, University of Science and Technology of China, Hefei 230031, China
[e] Department of Biomedical Engineering, The Ohio State University, Columbus, OH 43210, USA
[*] Corresponding Author: Shuwei Shen, E-mail: swshen@ustc.edu.cn, Peng Yao, E-mail: yaopeng@ustc.edu.cn, Ronald X. Xu, E-mail: xu.202@osu.edu

**Abstract**

Early and accurate diagnosis of Alzheimer's disease (AD) and its prodromal period mild cognitive impairment (MCI) is essential for the delayed disease progression and the improved quality of patients' life. The emerging computer-aided diagnostic methods that combine deep learning with structural magnetic resonance imaging (sMRI) have achieved encouraging results, but some of them are limit of issues such as data leakage and unexplainable diagnosis. In this research, we propose a novel end-to-end deep learning approach for automated diagnosis of AD and localization of important brain regions related to the disease from sMRI data. This approach is based on a two-dimensional (2D) single model strategy and has the following differences from the current approaches: 1) Convolutional Neural Network (CNN) models of different structures and capacities are evaluated systemically and the most suitable model is adopted for AD diagnosis; 2) a data augmentation strategy named Two-stage Random RandAugment (TRRA) is proposed to alleviate the overfitting issue caused by limited training data and to improve the classification performance in AD diagnosis; 3) an explainable method of Grad-CAM++ is introduced to generate the visually explainable heatmaps that localize and highlight the brain regions that our model focuses on and to make our model more transparent. Our approach has been evaluated on two publicly accessible datasets for two classification tasks of AD vs. cognitively normal (CN) and progressive MCI (pMCI) vs. stable MCI (sMCI). The experimental results indicate that our approach outperforms the state-of-the-art approaches, including those using multi-model and three-dimensional (3D) CNN methods. The resultant localization heatmaps from our approach also highlight the lateral ventricle and some disease-relevant regions of cortex, coincident with the commonly affected regions during the development of AD.

---

[1] Data used in preparation of this article were obtained from the Alzheimer's Disease Neuroimaging Initiative (ADNI) database (http://www.loni.ucla.edu/ADNI). As such, the investigators within the ADNI contributed to the design and implementation of ADNI and/or provided data but did not participate in analysis or writing of this report. ADNI investigators include (complete listing available at http://adni.loni.ucla.edu/wpcontent/uploads/how_to_apply/ADNI_Authorship_List.pdf).

[2] Data used in the preparation of this article was obtained from the Australian Imaging Biomarkers and Lifestyle flagship study of ageing (AIBL) funded by the Commonwealth Scientific and Industrial Research Organisation (CSIRO) which was made available at the ADNI database (www.loni.usc.edu/ADNI). The AIBL researchers contributed data but did not participate in analysis or writing of this report. AIBL researchers are listed at www.aibl.csiro.au.

## 1. Introduction

Alzheimer's disease (AD) is the most common type of dementia [1]. It is estimated that 131 million people worldwide will suffer from AD and other dementias by 2050, presenting a great healthcare challenge in the 21st century [2]. Mild cognitive impairment (MCI) represents a slight decline of mental ability along the continuum from normal cognition to AD, while over 33% of MCI subjects will progress to AD within five or more years [2, 3]. Currently, there is no curative treatment for AD. However, the progression of the disease can be slowed down through medications, exercise and memory training [4]. In this regard, Early detection of AD and accurate diagnosis of MCI are critical for delaying the disease progress and improving the patient's quality of life [5]. Since various neuroimaging tools, such as structural magnetic resonance imaging (sMRI) and positron emission tomography (PET), can differentiate neuropathological alterations associated with these diseases, they have been increasingly used for clinical diagnosis of AD and MCI [6].

In recent years, many researchers have developed computer-aided diagnostic systems by combining machine learning methods and sMRI data to identify the progression of AD [7-11]. Herein, the primary research tasks include the classification of AD versus cognitively normal (CN) [12] and the prediction of conversion from MCI toward AD (stable MCI (sMCI) versus progressive MCI (pMCI)) [4]. In these studies, the predefined features are first obtained from image preprocessing procedures, and then different types of classifiers are applied for classification tasks [8-10]. Since the feature selection and the classification algorithms are executed independently in traditional machine learning methods [13], this may lead to the potential loss of information associated with the classification tasks [14].

Deep learning is a state-of-the-art machine learning technique capable of extracting low-to-high level feature representations automatically from large and high-dimensional data sets, superior to the traditional machine learning methods [15]. As one of the most popular deep learning architectures, Convolutional Neural Network (CNN) has recently been explored for AD diagnosis [5, 16-19]. Lian et al. proposed a hierarchical fully convolutional network to construct the hierarchical classifier for AD diagnosis [19]. Liu et al. proposed a multi-model deep learning method for hippocampal segmentation and AD diagnosis [5]. Despite these encouraging results, the credibility of some studies in CNN-assisted AD diagnosis is hindered by data leakage issues [12]. Wen et al. analyzed the reasons that cause data leakage and pointed that a subject simultaneously appearing in training, validation and test sets may virtually increase the performance of the CNN models [12]. Backstrom et al. also verified that the diagnostic accuracy of the unbiased splitting (at the subject level) is 8% lower than that of the biased splitting (at the slice level) [20]. Two-dimensional (2D) CNN models, such as DenseNet [21] and EfficientNet [22], have been successfully implemented in natural image classification and are also explored in AD diagnosis [12]. 2D models pre-trained on ImageNet [23] are readily applicable to small-scale medical image datasets by transfer learning to achieve better performance [24]. In addition, many slices can be extracted from a single 3D image to increase the amount of training data in 2D models [12]. However, it was also reported that the AD classification accuracy for 2D CNN models is 10% lower than that of three-dimensional (3D) CNN models [12]. We will focus on 2D CNN models with the hypothesis that they will yield the classification accuracy comparable to a 3D model after algorithm optimization.

This research aims at addressing several unsolved problems associated with CNN-assisted AD diagnosis. First of all, there is no systematic comparison of the classification performance for different CNN models in AD diagnosis. For example, Wen et al. [12] and Valliani et al. [25] both used ResNet-18 in their studies but discarded other ResNet models [26]. Second, automated augmentation strategies have not been introduced in CNN-assisted AD diagnosis despite their demonstrated effectiveness in alleviating the overfitting issue caused by limited training data. Finally, many CNN models for AD diagnosis cannot provide the explanations of their predictions due to the "black box" nature of deep learning.

When performing classification tasks on large-scale image datasets, ameliorating model structure from initial AlexNet [27] to EfficientNet [22] or increasing the capacity of the similar model structures can always achieve better performance [22, 26]. However, this is not always correct on small-scale image datasets because the increased capacity may cause the model to transition from an under-fitting area to an over-fitting area [28]. Considering that even Alzheimer's Disease Neuroimaging Initiative (ADNI), one of the largest public datasets for AD diagnosis, has limited amount of data, the first question we focus on is: which model structure yields the best performance and what capacity of models in similar structures is most suitable for AD diagnosis? In this research, we try to identify the most suitable model by assessing the performance of CNNs with different structures and capacities.

At the same time, we need to further alleviate the overfitting issue caused by the limited amount of data. Data augmentation is one of the effective methods to alleviate the overfitting issue and finally improve the generalization of models. Since the conventional data augmentation strategies are problem specific, it is difficult to extend the same strategies to different applications and fields. Automated augmentation strategies are expected to overcome this shortcoming [29-31] and various automated augmentation strategies, such as AutoAugment [29] and RandAugment [31], have proven their effectiveness in alleviating overfitting and improving model robustness for natural image classification. Considering the difference between natural image datasets and sMRI datasets, direct use of data augmentation strategies developed for the former may not be the best choice. In this research, we propose a Two-stage Random RandAugment (TRRA) for improved classification performance in AD diagnosis.

Recently, visual explanations of CNN models on large-scale image dataset for enhanced transparency has attracted more and more research attention. Gradient-weighted Class Activation Mapping (Grad-CAM) introduces the gradients of the predicted target with respect to the final convolutional layer to generate a localization heatmap highlighting the areas that are important to the predicted target in the image [32]. As an improved version of Grad-CAM, Grad-CAM++ generates better visual explanations of model predictions to improve the model transparency [33]. To the best of our knowledge, no research has been reported yet to generate 2D visual explanations of MCI conversion prediction task despite some explorations of explanation heatmaps in AD classification task [34, 35].

In this paper, we propose a novel end-to-end deep learning approach based on a 2D single model for automated diagnosis of AD and localization of important brain regions related to the disease from the sMRI data. The main contributions of this research are summarized as follows:

1) CNN models of different structures and capacities are evaluated systemically and the most suitable model is adopted for AD diagnosis. To the best of our knowledge, this is the first report of using EfficientNet for AD diagnosis.

2) A TRRA data augmentation strategy is proposed to alleviate the overfitting issue caused by limited training data and to improve the classification performance in AD diagnosis.

3) An explainable method of Grad-CAM++ is introduced to generate the visually explainable heatmaps that localize and highlight the brain regions that our model focuses on and to make our model more transparent.

## 2. Methods

2.1 Participants and data preprocessing

Data used in this research were obtained from Alzheimer's Disease Neuroimaging Initiative (ADNI) database[3] and Australian Imaging Biomarkers and Lifestyle flagship study of ageing (AIBL) database[4]. ADNI dataset is one of the largest publicly accessible datasets used for AD diagnosis and has been widely used in scientific research. AIBL dataset has the similar inclusion criteria and image acquisition procedures with ADNI dataset and is commonly used to further evaluate the generalization ability of the models. The ADNI was launched in 2003 as a public-private partnership, led by Principal Investigator Michael W. Weiner, MD. The primary goal of ADNI has been to test whether serial MRI, PET, other biological markers, and clinical and neuropsychological assessment can be combined to measure the progression of MCI and early AD. For up-to-date information, see www.adni-info.org. Data in AIBL database was collected by the AIBL study group. AIBL study methodology has been reported previously [36]. Informed consent was acquired from all participants, and the ethics committee of the leading institution of each dataset approved their research. All the data used in this research are from baseline assessments. Table 1 and Table 2 summarize the demographics, the mini-mental state examination (MMSE) scores, and the global clinical dementia rating (CDR) scores of the ADNI and AIBL participants.

Table 1 Summary of participant demographics, MMSE and CDR scores at baseline for ADNI.

|       | Subjects | Age                       | Gender        | MMSE               | CDR                      |
|-------|----------|---------------------------|---------------|--------------------|--------------------------|
| AD    | 333      | 75.0 ± 7.8 [55.1, 90.9]   | 150 F / 183 M | 23.2 ± 2.1 [18, 27]| 0.5: 156; 1: 176; 2: 1   |
| CN    | 338      | 74.4 ± 5.7 [59.8, 89.6]   | 174 F / 164 M | 29.1 ± 1.1 [24, 30]| 0: 338                   |
| sMCI  | 296      | 72.2 ± 7.44 [55.0, 88.6]  | 119 F / 177 M | 28.0 ± 1.7 [23, 30]| 0.5: 296                 |
| pMCI  | 302      | 74.3 ± 7.1 [55.2, 91.7]   | 123 F / 179 M | 26.8 ± 1.9 [19, 30]| 0.5: 300; 1: 2           |

Values are presented as Means ± S.D. [range]. M: male, F: female

Table 2 Summary of participant demographics, MMSE and CDR scores at baseline for AIBL.

|    | Subjects | Age                      | Gender        | MMSE                | CDR                            |
|----|----------|--------------------------|---------------|---------------------|--------------------------------|
| AD | 77       | 75.0 ± 7.7 [55.5, 93.4]  | 43 F / 34 M   | 20.6 ± 5.3 [6, 29]  | 0.5: 29; 1: 40; 2: 6; 3: 2     |
| CN | 450      | 73.1 ± 6.2 [60.3, 92.1]  | 263 F / 187 M | 28.8 ± 1.2 [25, 30] | 0: 425; 0.5: 25                |

Values are presented as Means ± S.D. [range]. M: male, F: female

---

[3] http://adni.loni.usc.edu/
[4] https://aibl.csiro.au/

The ADNI and AIBL data are preprocessed following a step-by-step procedure. First, all the data are converted into the Brain Imaging Data Structure (BIDS) format [37]. Second, the N4ITK method is used for the bias field correction [38]. Third, the SyN algorithm [39] from ANTs [40] is used for affine registration that aligns each image to the MNI space with the ICBM 2009c nonlinear symmetric template [41, 42]. Finally, the registered images are cropped to remove the background, resulting in the images of size 169×208×179, with 1 mm isotropic voxels. For each subject, we obtain 129 slices of RGB images by discarding the first twenty and last twenty slices along the sagittal direction and copying each of the remaining slices to the R, G, and B channels. All the preprocessing procedures are performed using the Clinica [12, 43, 44] and the ANTs [45, 46] software packages.

2.2 Overview of the proposed deep learning approach

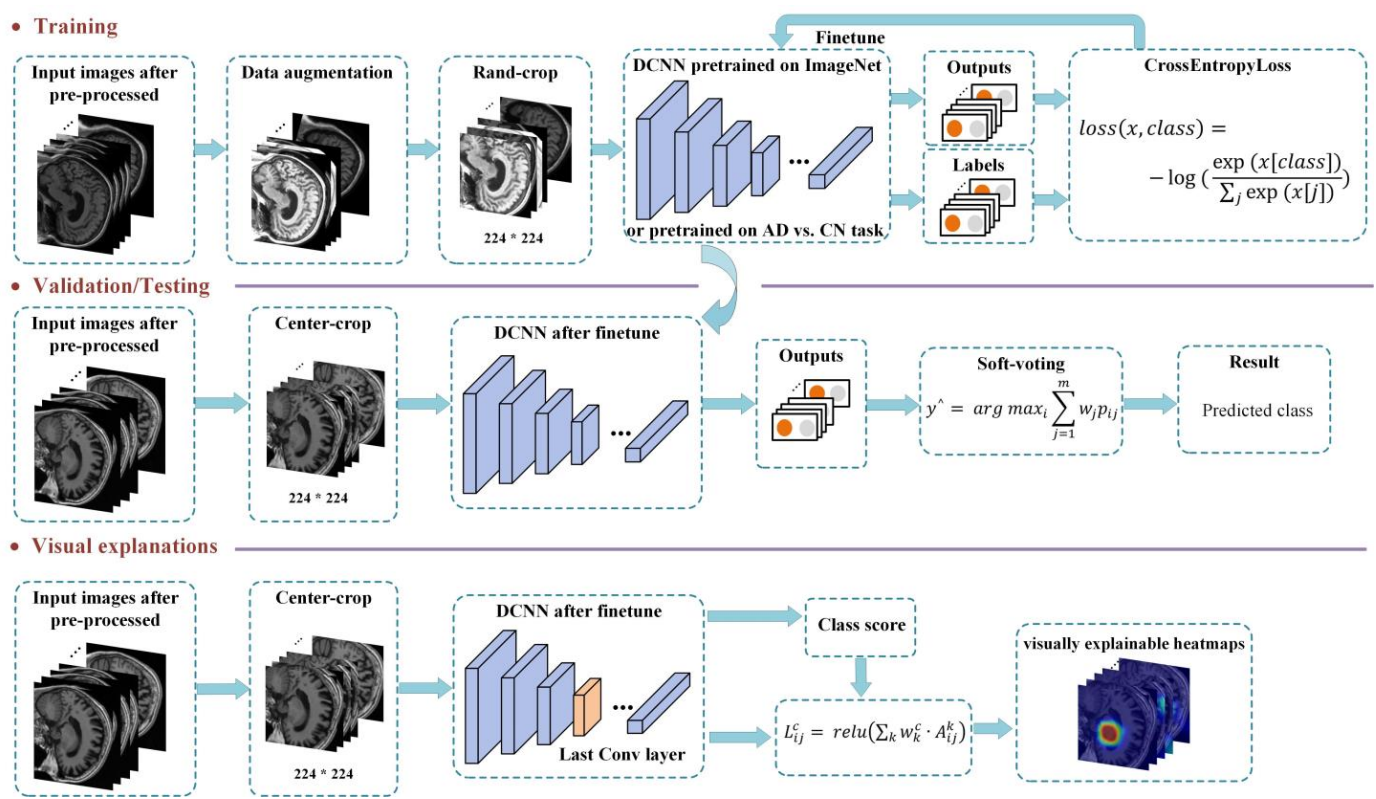

Fig. 1. The flowchart of the proposed deep learning approach.

Fig. 1 shows the flowchart of our proposed approach that includes the sequential stages of training, validation/testing and visual explanation. The pre-processed images are firstly resized from 208×179 to 297×256 in all the stages. During the training stage, the TRRA data augmentation strategy is applied to each image in the training set and the resultant image is randomly cropped to match the size of 224×224 required by the CNN models. For the AD classification task, we use the model pre-trained on the ImageNet dataset and fine-tune it on the ADNI training set. For the MCI conversion prediction task, we also investigate the possibility of transferring a CNN model pre-trained on AD classification task to this task. For each classification task, the model generates two prediction outputs per image and the cross-entropy loss function expressed as Equation1 is adopted:

$$loss(x, class) = -\log \left( \frac{\exp(x[class])}{\sum_j \exp(x[j])} \right) \quad (1)$$

where $class \in \{0,1\}$ specifies the ground-truth class and $x$ is the values predicted by the model. No data augmentation strategy is applied during the validation/test stage, and the input image is only center cropped to ensure the repeatability of each test. For each subject, soft voting is used to generate the subject-level decision [47]. First, SoftMax normalization is carried out on the output of all slices from the same patient to obtain the predicted probability $p$. Second, the number of correct predictions for the j-th slice of all subjects on the validation set is divided by all the number of correct predictions for 129 slices in order to obtain the weight of the j-th slice $w_j$. Finally, the subject-level decision is made based on the following formula:

$$y^\wedge = \arg \max_i \sum_{j=1}^{129} w_j p_{ij} \quad (2)$$

where $y^\wedge$ is the class of a subject that is finally predicted and $i \in \{0,1\}$ contains all the possible classes. For the AD classification task, the subject will be predicted as AD (CN) if $y^\wedge = i = 1\ (0)$. For MCI conversion prediction task, the subject will be predicted as pMCI (sMCI) if $y^\wedge = i = 1\ (0)$. The weight $w_j$ reflects the importance of each slice and $w_j$ calculated on the validation set will be retained and used when evaluating on the test set [12].

For the visual explanation stage, the gradient weights $\alpha_{kc}^{ij}$ for the predicted class $c$ and the feature map $A^k$ is firstly calculated using the following formula:

$$\alpha_{kc}^{ij} = \frac{\frac{\partial^2 Y^c}{(\partial A_{ij}^k)^2}}{2\frac{\partial^2 Y^c}{(\partial A_{ij}^k)^2} + \sum_a \sum_b A_{ab}^k \left\{ \frac{\partial^3 Y^c}{(\partial A_{ij}^k)^3} \right\}} \quad (3)$$

where $Y^c$ is the predicted class score, and $A^k$ is the k-th feature map of the last convolutional layer. $(i,j)$ and $(a,b)$ are the position of the feature map $A^k$. Then, the gradient of $Y^c$ with respect to the position $(i,j)$ of the feature map $A^k$ is calculated. Then, the weights $w_k^c$ is calculated as:

$$w_k^c = \sum_i \sum_j \alpha_{kc}^{ij} \cdot relu \left( \frac{\partial Y^c}{\partial A_{ij}^k} \right) \quad (4)$$

where $relu$ function is used to get positive gradients. Finally, the visually explainable heatmap is generated by combining the weights $w_k^c$ and all $K$ feature maps:

$$L_{ij}^c = relu(\textstyle\sum_k w_k^c \cdot A_{ij}^k) \qquad (5)$$

## 2.3 Convolutional Neural Network (CNN) models

Five different CNN structures from classic VGG series [48] to the latest EfficientNet series [22] are adopted in this research and their detailed information is list in Table3. For all the models, the last fully connected (FC) layer is replaced with a new FC layer with 2 output nodes.

Table 3 Detailed information of CNNs with different structures and different parameters.

| Model | Params(M) | FLOPs(B) | Model | Params(M) | FLOPs(B) | Model | Params(M) | FLOPs(B) |
|---|---|---|---|---|---|---|---|---|
| VGG-11 | 132.9 | 7.6 | SE-ResNet-50 | 28.1 | 3.9 | EfficientNet-B1 | 7.8 | 0.7 |
| VGG-13 | 133.1 | 11.3 | SE-ResNet-101 | 49.3 | 7.6 | EfficientNet-B2 | 9.1 | 1.0 |
| VGG-16 | 138.4 | 15.5 | SE-ResNet-152 | 66.8 | 11.4 | EfficientNet-B3 | 12.2 | 1.8 |
| VGG-19 | 143.7 | 19.7 | SENet-154 | 115.1 | 20.8 | EfficientNet-B4 | 19.3 | 4.2 |
| ResNet-18 | 11.7 | 1.8 | DenseNet-121 | 8.0 | 2.9 | EfficientNet-B5 | 30.4 | 9.9 |
| ResNet-34 | 21.8 | 3.7 | DenseNet-169 | 14.2 | 3.4 | EfficientNet-B6 | 43.0 | 19 |
| ResNet-50 | 25.6 | 4.1 | DenseNet-201 | 20.0 | 4.4 | EfficientNet-B7 | 66.4 | 37 |
| ResNet-101 | 44.6 | 7.9 | DenseNet-161 | 28.7 | 7.8 | | | |
| ResNet-152 | 60.2 | 11.6 | EfficientNet-B0 | 5.3 | 0.39 | | | |

VGG proves that change in depth of the network will affect the final performance in the large-scale image recognition task [48], but excessive increasing of the network depth will degrade their performance. ResNet alleviates this problem by introducing residual learning [26]. They add a skip-connection that bypasses the non-linear transformations with an identity function:

$$y = F(x, \{W_i\}) + x. \qquad (6)$$

where $x$ and $y$ are the input and output vectors of the layer, and $F(x, \{W_i\})$ represents the non-linear transformation. Unlike the method of ResNet, DenseNet achieves similar effect through concatenating features [21]. Consequently, the $l^{th}$ layer receives the feature-maps of all preceding layers as follows:

$$x_l = H_l([x_0, x_1, \dots, x_{l-1}]) \qquad (7)$$

where $[x_0, x_1, \dots, x_{l-1}]$ refers to the concatenation of the feature-maps produced in layers $0, \dots, l-1$. $H_l(\cdot)$ is a composite function of three consecutive operations: batch normalization (BN), a rectified linear unit (ReLU) and a 3×3 convolution.

Tan et al. systematically studies the principle of model scaling and they believe that balancing network depth, width and image resolution in the process of scaling the model can bring better performance [22]. The new compound scaling method they propose uses a compound coefficient $\emptyset$ to uniformly scales network width ($w$), depth ($d$), and resolution ($r$) in a principled way:

$$d = \alpha^{\emptyset}; w = \beta^{\emptyset}; r = \gamma^{\emptyset} \qquad (8)$$
$$s.t. \quad \alpha \cdot \beta^2 \cdot \gamma^2 \approx 2 \quad \alpha \geq 1, \beta \geq 1, \gamma \geq 1$$

Intuitively speaking, $\emptyset$ is given by the user to control how many resources are available for model scaling, and $\alpha, \beta, \gamma$ specifies how to allocate these additional resources to the network depth, width, and resolution respectively. They used different composite coefficients $\emptyset$ in Equation 8 to scale the baseline EfficientNet-B0 to obtain a total of 7 different EfficientNet models. The main building block used in EfficientNet named mobile inverted bottleneck [49, 50] is shown in Fig. 2.

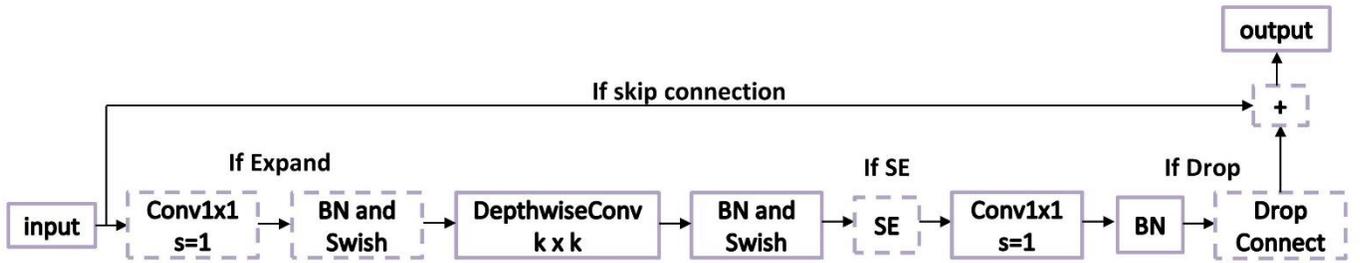

Fig. 2. Schematic diagram of mobile inverted bottleneck used in EfficientNet.

2.4 Data augmentation strategy

RandAugment (RA) proposed by Cubuk et al. [31] is the most advanced data augmentation strategy so far that significantly improve classification performance and model robustness for image classification. The search space of RA consists of $K$ ($K = 16$) available transformations. RA contains two interpretable integer hyperparameters $N$ and $M$. $N$ is the number of transformations applied to a training image sequentially, and $M$ is the magnitude for all the transformations. When performing data augmentation, RA randomly select $N$ transformations in the search space and apply them to the image according to the preset magnitude $M$. Given $N$ transformations for a training image, RA may thus express $K^N$ potential policies.

Inspired by RA, we propose a novel data augmentation strategy called Two-stage Random RandAugment (TRRA) which is more suitable for AD diagnosis compared with RA. Specifically, we first expand the types of transformations available in the search space from 16 to 23 and all available transformations and corresponding range of magnitude are listed in Table 4 (7 newly added transformations are bolded). Then, the 23 transformations are further divided into two categories of [$color$] and [$shape$]. Then, the magnitude $M$ is upgraded from a fixed integer to an integer randomly sampled between the preset two values. Finally, a probability parameter $P$ is introduced to control whether the selected transformation should be executed or not so that each transformation has the probability of $1 - P$ to remain the input image unchanged.

The rationale of TRRA design lies on the following three aspects. First, we think that adding 7 kinds of transformations and setting $M$ randomly sampling between two values can further increase the diversity and quantity of training data. Second, we believe that color attributes related transformations and shape attributes related transformations contribute differently to the classification performance. In RA, each operation is randomly selected from all the transformations without differentiating categories. Therefore, it is likely that most of the operations are selected from the category with relatively small contributions in the case of $N > 1$. So, we use two hyperparameters $N_{color}$ and $N_{shape}$ to explicitly specify the number of transformations selected from the two categories. Finally, we believe that superimposing too many transformations on the input image will destroy its inherent characteristics despite the increased diversity of training data by data augmentation. The probability parameter $P$ and $N_{color(shape)}$ can limit the data augmentation process to a suitable range. So, we introduce third hyperparameter $P$ to control the probability of execution of each operation.

Table 4 List of all transformations can be selected during the search using TRRA.

| | Operation Name | Description | Range of magnitude |
|---|---|---|---|
| [color] | Auto Contrast | Maximize (normalize) image contrast. | - |
| | Equalize | Equalize the image histogram. | - |
| | Invert | Invert (negate) the image. | - |
| | Posterize | Reduce the number of bits for each color channel. | [0, 4] |
| | Solarize | Invert all pixel values above a threshold. | [0, 256] |
| | Solarize Add | Add a value to the image and do solarize. | [0, 100] |
| | Color | Adjust image color balance. | [0.1, 1.9] |
| | Contrast | Adjust image contrast. | [0.1, 1.9] |
| | Brightness | Adjust image brightness. | [0.1, 1.9] |
| | Sharpness | Adjust image sharpness. | [0.1, 1.9] |
| | **Random noise** | Add a noise randomly sampled from a uniform distribution. | [0, 0.4] |
| | **Gaussian noise** | Add a noise randomly sampled from the Gaussian distribution. | [0, 0.4] |
| | **Gaussian blur** | Gaussian blur filter. | [0, 2.0] |
| [shape] | **Horizontal flip** | Flip the image Horizontally (left to right). | - |
| | **Vertical flip** | Flip the image vertically (top to bottom). | - |
| | Rotate | Rotate the image according to magnitude. | [0, 30] |
| | Shear X | Shear the image along the horizontal axis. | [0, 0.3] |
| | Shear Y | Shear the image along the vertical axis. | [0, 0.3] |
| | Cutout | Set a random square patch with a side length of magnitude, pixels inside turn gray. | [0, 40] |
| | Translate X | Move the image along the horizontal axis. | [0, 100] |
| | Translate Y | Move the image along the vertical axis. | [0, 100] |
| | **Scale** | Scale the image horizontally and vertically with equal magnitude degrees. | [0.9, 1.4] |
| | **Scale XY** | Scale the image horizontally and vertically with different magnitude degrees. | [0.9, 1.4] |

The following ablation experiments are designed to verify the contribution of each improvement of TRRA to classification performance.

1) To investigate the contribution of 7 newly added transformations: We expand the search space of RA by adding the 7 kinds of transformations so we can get RandAugment-23 (RA-23). RA-23 uses a fixed magnitude $M$ as RA.
2) To investigate the contribution of a random $M$: We change the magnitude $M$ of RA-23 from a fixed value to an integer randomly sampled between $[5, X]$ ($X \in [10, 30]$) to get Random-RandAugment-23 (RRA-23).
3) To investigate the contribution of dividing all transformations into two categories of $[color]$ and $[shape]$, we set the probability parameter $P$ in TRRA to 1, and then compare TRRA with RRA-23.
4) To investigate the contribution of the probability parameter $P$, we compare the performance of TRRA under different probability parameter $P$.

Fig. 3 shows the workflow of different data augmentation strategies. For RA, RA-23 and RRA-23, we perform a grid search to get their optimal performance. Specifically, hyperparameter $N$ is sampled from 1 to 8 in a step size of 1 for each strategy. Hyperparameter $M$ is sampled from 5 to 30 in a step size of 5 for RA and RA-23. For RRA-23, $M$ is an integer value randomly sampled between $[5, X]$, and $X$ is sampled from 10 to 30 in a step size of 5.

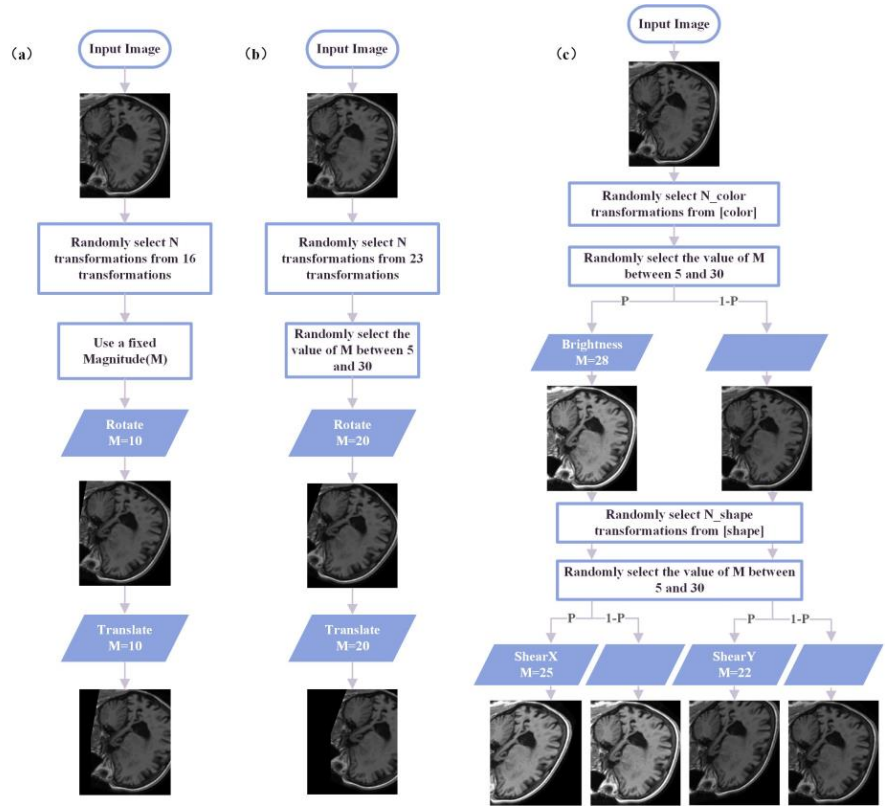

Fig. 3. Workflow of different data augmentation strategies. (a) RA and RA-23. (b) RRA-23. (c) TRRA.

2.5 Evaluation Metrics

The following commonly used metrics are chosen to evaluate the classification performance for AD diagnosis [19]: accuracy, sensitivity, specificity, and area under the receiver operating characteristic curve (AUC), where accuracy is used as the main evaluation metric.

2.6 Implementation

The performances of the proposed approach are evaluated using two binary tasks of AD classification (AD vs. CN) and MCI conversion prediction (sMCI vs. pMCI). The AD classification task is used as a baseline for evaluating the performance of different models and data augmentation strategies, and the best model is used for the MCI conversion prediction task.

To avoid data leakage, we adopt a previously reported method [12] to split the ADNI dataset and carefully check the results. Specifically, the ADNI dataset is split into the training/validation/test sets at the subject-level. The training and the validation sets are used for the selection of the model capacity of the five structures and the grid search of the hyperparameters of four data augmentation strategies. The test set only tests the best-performing model of each structure and the best hyperparameter combination of each data augmentation strategy. We ensure that age and sex distributions between training, validation and test sets are not significantly different. To avoid the influence caused by a single split, we carry out a total of three splits following the same ratio of training/validation/test sets (6:2:2) as Backstrom et al [20]. All experiments are performed using these three splits so that the mean and standard deviation of the metrics can be obtained .

All the training and the testing tasks are performed on 2 NVIDIA GeForce GTX 2080Ti graphics cards using Pytorch. To prevent overfitting, we adopt an early stopping strategy: when the validation accuracy doesn't improve for a continuous 20 epochs, the training process will stop, otherwise, the training will continue to the end of the predefined periods. The selected model is the one which obtain the highest validation accuracy during training. Batch size for model training in this study is 128, but it is reduced for some of the large models to match the memory capacity of the graphic cards.

3. Results

3.1 Comparison study of different CNN models

In this part of experiments, we first compare different CNN models on ADNI validation set to determine the best-performing model of each structure, and then test them on ADNI test set. The detailed experimental results of different CNN models on ADNI validation set can be found in Appendix A.

The AD classification performance on ADNI test set of CNN models in different structures is presented in Table 5. ResNet-18 is also selected for comparing with the results in the literature [12, 25]. As the data in Table 5 show: 1) Accuracy of ResNet-18 without applying data augmentation is 0.774. This is very similar to Wen et al. (0.760) [12] and Valliani et al. (0.788) [25], which indicates no data leakage in our evaluation. 2) The models of different CNN structures in Table 5 are all in the moderate capacity rather than the maximum capacity, which indicates that the models with moderate capacity instead of maximum capacity achieve the best performance. 3) The classification performance of each model applying TRRA show similar significant improvement. The general improvement of more than 10% in them indicate that TRRA is universal to different CNN structures. 4) EfficientNet-B1 and DenseNet-169 both achieve the highest accuracy (0.932) on the ADNI test set. Combining the above observations and data in Appendix A, we can see that more advanced model structures can achieve better performance, and models in similar structure with moderate capacity rather than the largest one can achieve better performance. Considering that EfficientNet-B1 has the highest accuracy on both ADNI validation set and ADNI test set, it is used in the following experiments.

Table 5 AD classification performance on ADNI test set of best-performing model of each CNN structure.

| Model | Performance with TRRA | | | | Performance without data augmentation | | | |
| --- | --- | --- | --- | --- | --- | --- | --- | --- |
| | Accuracy | Sensitivity | Specificity | AUC | Accuracy | Sensitivity | Specificity | AUC |
| VGG-13 | 0.912±0.009 | 0.904±0.026 | 0.920±0.043 | 0.962±0.009 | 0.789±0.018 | 0.788±0.062 | 0.791±0.024 | 0.872±0.022 |
| ResNet-18 | 0.912±0.004 | 0.874±0.007 | 0.950±0.007 | 0.957±0.008 | 0.774±0.025 | 0.753±0.014 | 0.796±0.037 | 0.853±0.030 |
| ResNet-50 | 0.920±0.014 | 0.904±0.031 | 0.935±0.019 | 0.961±0.011 | 0.784±0.014 | 0.727±0.025 | 0.841±0.051 | 0.865±0.027 |
| SE-ResNet-101 | 0.922±0.015 | 0.889±0.019 | 0.955±0.012 | 0.960±0.009 | 0.794±0.014 | 0.722±0.038 | 0.866±0.064 | 0.875±0.029 |
| DenseNet-169 | 0.932±0.006 | 0.904±0.014 | 0.960±0.019 | 0.961±0.009 | 0.800±0.018 | 0.778±0.040 | 0.821±0.074 | 0.869±0.026 |
| EfficientNet-B1 | **0.932±0.006** | **0.924±0.000** | **0.940±0.012** | **0.961±0.012** | 0.777±0.019 | 0.692±0.038 | 0.861±0.070 | 0.870±0.027 |

Values are presented as Means ± S.D. Accuracy with DA: AD vs. CN classification using TRRA, Accuracy without DA: AD vs. CN classification without applying data augmentation.

3.2 Comparison study of different data augmentation strategies

In this part of experiments, we first perform a grid search on ADNI validation set to determine the optimal hyperparameter combination of each data augmentation strategy, and then test them on ADNI test set. The detailed experimental results of EfficientNet-B1 with different data augmentation strategies on ADNI validation set can be found in Appendix B.

The AD classification performance on ADNI test set of each data augmentation strategy is presented in Fig. 4. Observations from Fig. 4 show that: 1) RA-23 performs better than RA, which indicates that adding 7 kinds of transformations in the search space helps to improve classification performance. 2) RRA-23 performs better than RA-23, which indicates that compared with the fixed magnitude, a magnitude randomly sampled between two values helps to improve classification performance.

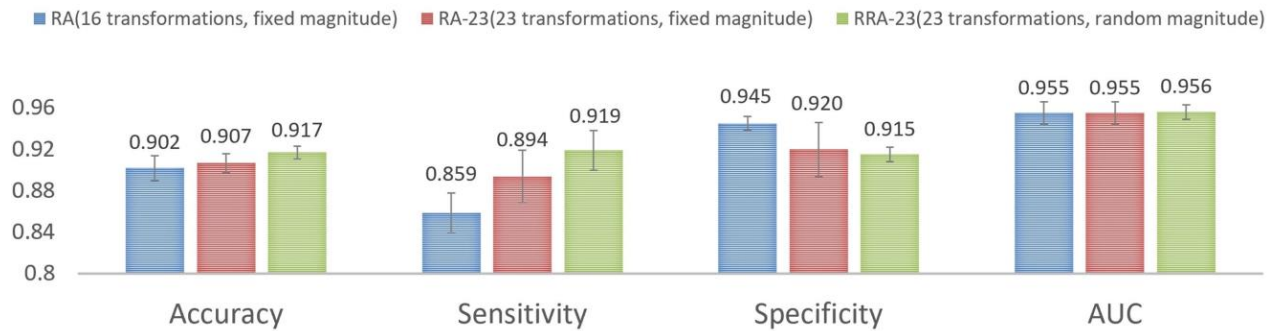

Fig. 4. AD classification performance on ADNI test set of the of RA, RA-23, and RRA-23 in optimal hyperparameters.

As shown in Fig. 4, RRA-23 helps to get the best accuracy of 0.917 on ADNI test set when the hyperparameters $N$ is 7 and $M$ is randomly sampled from [5, 30]. So, we set the sum of $N_{color}$ and $N_{shape}$ as 7, and $M$ randomly sample between [5, 30] in searching the optimal hyperparameters for TRRA. The detailed experimental results of EfficientNet-B1 with TRRA under different hyperparameters on ADNI validation set refer to Appendix C. From the results we observe that randomly selecting five transformations from the [$color$] category and randomly selecting two transformations from the [$shape$] category achieve the best performance, which indicates that [$color$] category contributes more to classification performance than [$shape$] category.

The AD classification performance on ADNI test set of TRRA is presented in Fig. 5. Observations from Fig. 4 and Fig. 5 show that: 1) TRRA performs better than RRA-23. The accuracy of TRRA is 0.930 when $P$ is equal to 1 and is further improved compared with 0.917 achieved by RRA-23, which indicates dividing all transformations into two categories of [$color$] and [$shape$] is better for classification performance. 2) The accuracy and AUC of $P$ is 0.9 are improved by 0.2% and 0.3% compared that when $P$ is 1, which indicates that $P$ can help improve classification performance.

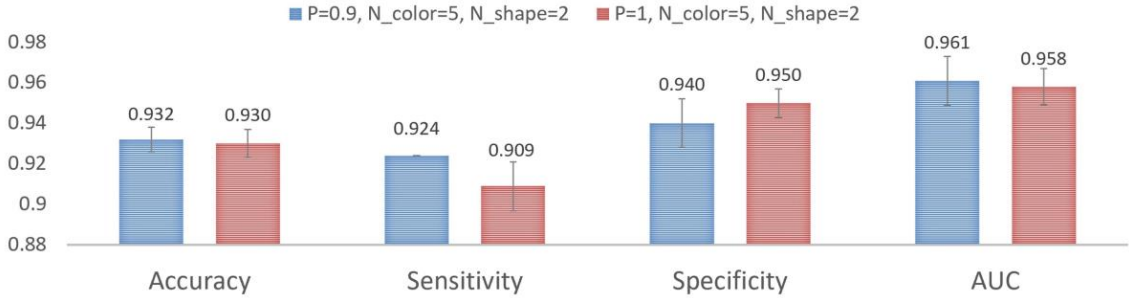

Fig. 5. AD classification performance on ADNI test set of TRRA.

3.3 Classification Performance on MCI conversion prediction task.

Fig. 6 shows the performance of MCI conversion prediction task using different pre-training methods and data augmentation strategies. We first train ImageNet pre-trained EfficientNet-B1 on ADNI training set of MCI conversion prediction task without using data augmentation and achieve accuracy of 0.700 on test set. Then, we use the EfficientNet-B1 model performing best on AD classification task, and fine-tune it without data augmentation on the ADNI training set of MCI conversion prediction task, and accuracy on ADNI test set achieves 0.751. Compared with the ImageNet pre-trained model, using AD classification task for pre-training improves the accuracy of the MCI conversion prediction task by 5.1%. This proves the effectiveness of using AD classification task for pre-training. Finally, we use TRRA to perform data augmentation during the training process on MCI conversion prediction task, accuracy on ADNI test set is further improved to 0.829, which is increased by 7.8% in the comparison with no data augmentation.

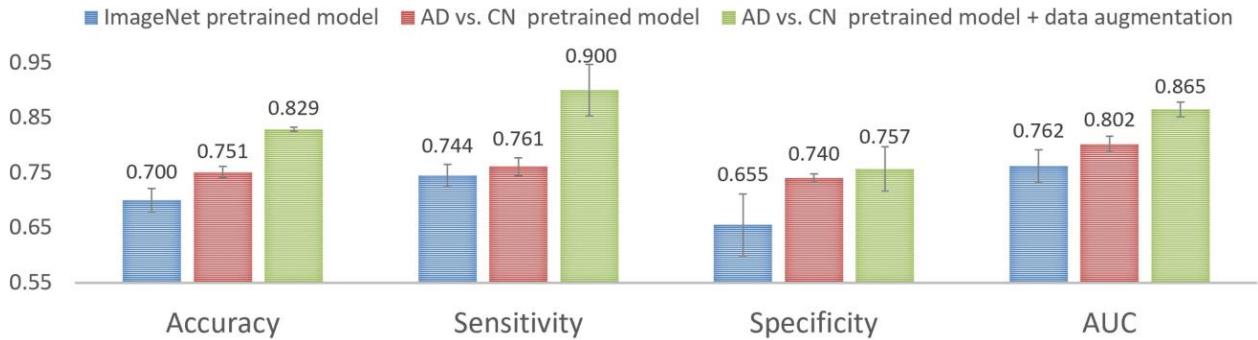

Fig. 6. The classification performance on MCI conversion prediction task.

3.4 Classification Performance on AIBL dataset

AIBL dataset is used to further evaluate the generalization of our proposed approach. Specifically, we choose the EfficientNet-B1 models performing best on ADNI dataset (accuracy is 0.932±0.006 on ADNI test set) and take the AIBL dataset containing 77 AD subjects and 450 CN subjects for testing. Accuracy on AIBL dataset is 0.920±0.006. Noticeably, we do not further fine-tune the model on AIBL dataset, and use all the data as a test set, which is a more difficult choice. To the best of our knowledge, only Wen et al. [12] test the ADNI trained model on the AIBL dataset, and our performance is better than theirs. The results verify that our approach generalizes well not only on the dataset from the same research, but also on the dataset from a similar study. Table 6 presents the details experimental results.

Table 6 AD vs. CN classification performance on AIBL dataset.

| Approach | Accuracy | Sensitivity | Specificity | AUC |
|---|---|---|---|---|
| Our | 0.920±0.006 | 0.818±0.021 | 0.937±0.004 | 0.939±0.003 |
| Wen et al. | 0.896±0.011 | 0.771±0.051 | 0.918±0.020 | - |

Values are presented as Means ± S.D.

3.5 Visually explainable heatmaps

Grad-CAM++ has been previously introduced to generate visually explainable heatmaps helping to localize and highlight the brain regions related with predicted target. We only apply Grad-CAM++ to the best-performing models on the two classification tasks. The visually explainable heatmaps indicate that our models focus on the lateral ventricle and some regions of cortex, as presented in Fig. 7. In AD classification task, for AD subjects, model pays more attention to dilation of the lateral ventricle and cortical atrophy (atrophic temporal lobe, insular lobe, frontal lobe). It is interesting to notice that in the MCI conversion prediction task, for pMCI subjects, the regions that the model focuses on are similar with AD subjects, and for sMCI subjects, the model focuses on the parietal and frontal lobes.

The dilation of the lateral ventricle and cortical atrophy are the macroscopic features of neuropathological alterations in AD and can be recognized by MRI in the early stages of AD development [51, 52]. The focus regions of our models are consistent with these regions, which brings good classification performance.

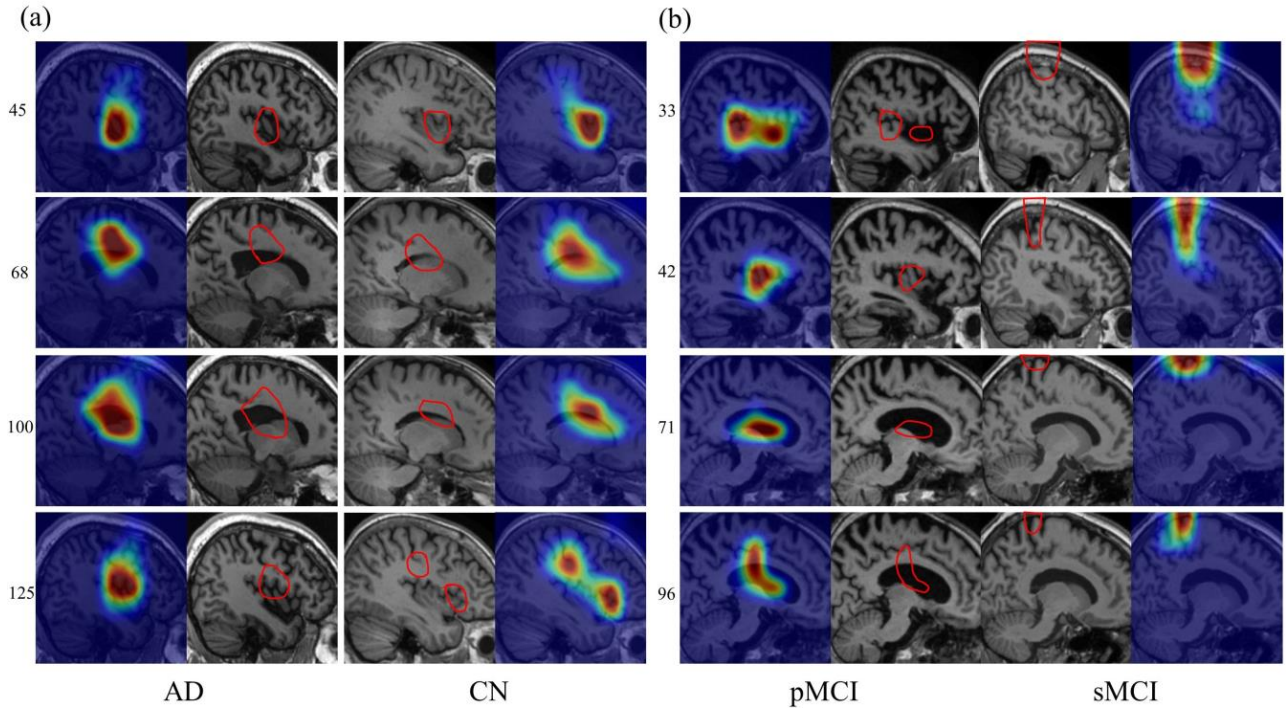

Fig. 7. The visual explanation results on AD classification and MCI conversion prediction tasks. The highlighted regions on heatmaps are of higher correlation with the predicted class and the boundary of the most important red area is drawn on the original image for easy observation. The numbers on the left indicate the slice positions.

3.6 Comparison with other methods

In this part, we provide a performance comparison table to further compare with most recent and state-of-the-art methods reported in the literature. Table 7 summarizes the methods using sMRI data from the ADNI dataset for AD diagnosis (no data leakage in all methods). As the data indicates, we rank first in accuracy and AUC on both classification tasks. The performance of the proposed 2D single model approach on the two classification tasks both outperforms the existing state-of-the-art methods including those using multi-model and 3D CNN.

Table 7 A comparative table of methodologies on both AD vs. CN task and pMCI vs. sMCI task using structural MRI data from the ADNI dataset.

| Study | AD vs. CN | | | | pMCI vs. sMCI | | | | Approach |
|---|---|---|---|---|---|---|---|---|---|
| | Accuracy | Sensitivity | Specificity | AUC | Accuracy | Sensitivity | Specificity | AUC | |
| Aderghal et al., 2018 [53] | 0.90 | 0.94 | 0.86 | | - | | | | 3D ROI-based |
| Bäckström et al., 2018 [20] | 0.90 | | | | - | | | | 3D subject-level |
| Cheng et al., 2017 [54] | 0.87 | 0.86 | 0.86 | 0.92 | - | | | | 3D patch-level |
| Cheng and Liu, 2017 [55] | 0.86 | 0.84 | 0.90 | 0.91 | - | | | | 3D subject-level |
| Li et al., 2017 [56] | 0.88 | 0.91 | 0.84 | 0.93 | - | | | | 3D subject-level |
| Li et al., 2018 [57] | 0.90 | 0.88 | 0.91 | 0.92 | - | | | | 3D patch-level |
| Lian et al., 2018 [19] | 0.90 | 0.82 | 0.97 | 0.95 | 0.81 | 0.53 | 0.85 | 0.78 | 3D patch-level |
| Mingxia Liu et al., 2018a [58] | 0.91 | 0.88 | 0.94 | 0.96 | 0.78 | 0.42 | 0.82 | 0.78 | 3D patch-level |
| Mingxia Liu et al., 2018b [59] | 0.91 | 0.87 | 0.93 | 0.96 | - | | | | 3D patch-level |
| Shmulev et al., 2018 [60] | - | | | | 0.62 | 0.75 | 0.54 | 0.70 | 3D subject-level |
| Valliani and Soni, 2017 [25] | 0.81 | | | | - | | | | 2D slice-level |
| Spasov et al., 2019 [61] | - | | | | 0.72 | 0.63 | 0.81 | 0.79 | 3D subject-level |
| Manhua Liu et al., 2020 [5] | 0.89 | 0.87 | 0.91 | 0.93 | - | | | | 3D ROI-based |
| Wen et al., 2020 [12] | 0.89 | 0.87 | 0.90 | | 0.74 | 0.80 | 0.68 | | 3D ROI-based |
| **Our method** | **0.93** | 0.92 | 0.94 | **0.96** | **0.83** | **0.90** | 0.76 | **0.87** | **2D slice-level** |

## 4. Discussion

As introduced previously, despite the existing research is encouraging, deep learning based diagnostic methods for AD and its prodromal period MCI still have some limitations. In this research, our proposed novel end-to-end deep learning approach based on a 2D single model can not only perform automated diagnosis of AD, but also generate visually explainable heatmaps helpful to localize the disease-related brain regions from sMRI data. The proposed approach outperforms the state-of-the-art approaches, including those using multi-model and 3D CNN methods.

For the AD diagnosis, our approach achieves the accuracy of 0.93, 0.83 for AD classification, MCI conversion prediction on the ADNI dataset respectively, and achieves an accuracy of 0.92 for AD classification on the AIBL dataset. For the first time, we systematically assessed CNN models with different structures and capacities for AD diagnosis. The results in Table 5 and Table A.1 indicate that more advanced model structures like EfficientNet and DenseNet can achieve better performance, and models in similar structure with moderate capacity rather than the largest one can achieve better performance.

Limited by lack of large-scale sMRI dataset, it is not easy to train a model of good classification performance for AD diagnosis. To alleviate this problem, we propose TRRA which is more suitable for AD diagnosis task than RA. The results of the ablation experiments in Fig. 4 and Fig. 5 presents the contribution of each improved elements of TRRA to classification performance. In addition, the experimental results in Fig. 6 also proves that pre-training on AD classification task can improve the classification performance of the MCI conversion prediction task.

Unbiased evaluation of performance is an essential task of deep learning, and the test set should not be used for hyperparameter selection. We therefore choose a rigorous evaluation strategy: Training and validation sets are used for the selection of the model capacity of the five structures and the grid search of the hyperparameters of four data augmentation strategies, and the test set is only adopted for evaluation of the final classifier.

Meanwhile, we introduce Grad-CAM++ to understand how the model makes the classification decision.

The heatmaps in Fig. 7 show that our approach pays more attention to the lateral ventricle and some regions of cortex, which has been proven to be affected during the development of AD. To the best of our knowledge, this is the first time to generate 2D visual explanation for MCI conversion prediction task.

Our approach greatly improves the classification performance of 2D CNN for AD diagnosis and the increases transparency of the model. The systematic evaluation of various CNN models provides a reference for subsequent studies. The proposed data augmentation strategy can greatly improve the diagnostic performance by alleviating the overfitting problem caused by the limited data in medical datasets, and it is also flexible to expend in other imaging modalities and medical datasets. Considering the potential scarcity of data in the medical field, we only use less invasive and cheaper sMRI data that can be obtained in non-tertiary medical center and medium hospitals, which can make our method applicable to a wider clinical environment.

## 5. Conclusion

In this research, we propose a novel end-to-end deep learning approach based on a 2D single model for automated diagnosis of AD and localization of important brain regions related to the disease from sMRI data. First, CNN models of different structures and capacities are evaluated systemically and the most suitable model is adopted for AD diagnosis. Then, a data augmentation strategy called TRRA able to alleviate overfitting is proposed to improve classification performance. Meanwhile, to understand how the model makes decisions and increase transparency of our approach, Grad-CAM++ is introduced to generate visually explainable heatmaps that localize and highlight the brain regions that our model focuses on. The effectiveness of our proposed approach has been extensively evaluated on two publicly accessible datasets. The experimental results indicate that our approach outperforms the state-of-the-art approaches, including those using multi-model and three-dimensional (3D) CNN methods. The resultant localization heatmaps from our approach also highlight the lateral ventricle and some disease-relevant regions of cortex, which is coincident with the commonly affected regions during the development of AD.

**Credit authorship contribution statement**

**Fan Zhang**: Methodology, Software, Validation, Formal analysis, Data Curation, Writing - Original Draft, Writing-Review & Editing. **Bo Pan**: Conceptualization, Formal analysis, Visualization. **Pengfei Shao**: Conceptualization, Methodology, Resources. **Peng Liu:** Methodology, Formal analysis. **Shuwei Shen**: Conceptualization, Methodology, Formal analysis, Writing-Review & Editing. **Peng Yao**: Methodology, Software, Formal analysis, Writing - Review & Editing, Supervision, Project administration. **Ronald X. Xu**:

Conceptualization, Methodology, Formal analysis, Writing - Review & Editing, Visualization, Supervision, Project administration.

**Declaration of interest**

The authors declare that there are no conflicts of interest.

**Acknowledgements**

Data collection and sharing for this project was funded by the Alzheimer's Disease Neuroimaging Initiative (ADNI) (National Institutes of Health Grant U01 AG024904) and DOD ADNI (Department of Defense award number W81XWH-12-2-0012). ADNI is funded by the National Institute on Aging, the National Institute of Biomedical Imaging and Bioengineering, and through generous contributions from the following: AbbVie, Alzheimer's Association; Alzheimer's Drug Discovery Foundation; Araclon Biotech; BioClinica, Inc.; Biogen; Bristol-Myers Squibb Company; CereSpir, Inc.; Cogstate; Eisai Inc.; Elan Pharmaceuticals, Inc.; Eli Lilly and Company; EuroImmun; F. Hoffmann-La Roche Ltd and its affiliated company Genentech, Inc.; Fujirebio; GE Healthcare; IXICO Ltd.; Janssen Alzheimer Immunotherapy Research & Development, LLC.; Johnson & Johnson Pharmaceutical Research & Development LLC.; Lumosity; Lundbeck; Merck & Co., Inc.; Meso Scale Diagnostics, LLC.; NeuroRx Research; Neurotrack Technologies; Novartis Pharmaceuticals Corporation; Pfizer Inc.; Piramal Imaging; Servier; Takeda Pharmaceutical Company; and Transition Therapeutics. The Canadian Institutes of Health Research is providing funds to support ADNI clinical sites in Canada. Private sector contributions are facilitated by the Foundation for the National Institutes of Health (www.fnih.org). The grantee organization is the Northern California Institute for Research and Education, and the study is coordinated by the Alzheimer's Therapeutic Research Institute at the University of Southern California. ADNI data are disseminated by the Laboratory for Neuro Imaging at the University of Southern California.

**Appendix A. Classification performance of different CNN models**

Table A.1 shows AD classification accuracy of different CNN models on ADNI validation set. We observe that regardless of whether TRRA is implemented during the training process, accuracy of models in similar structure increases to the maximum value and then decreases as the model capacity increases. This indicates that the models with moderate capacity instead of maximum capacity can achieve the best performance.

Table A.1 AD classification performance of different CNN models on ADNI validation set.

| Model | Accuracy without DA | Accuracy with DA | Model | Accuracy without DA | Accuracy with DA | Model | Accuracy without DA | Accuracy with DA |
|---|---|---|---|---|---|---|---|---|
| VGG-11 | 0.797±0.028 | 0.900±0.009 | SE-ResNet-50 | 0.762±0.019 | 0.902±0.012 | EfficientNet-B1 | **0.797±0.016** | **0.915±0.018** |
| VGG-13 | **0.805±0.025** | **0.907±0.013** | SE-ResNet-101 | **0.792±0.035** | **0.910±0.018** | EfficientNet-B2 | 0.764±0.038 | 0.912±0.015 |
| VGG-16 | 0.792±0.013 | 0.902±0.012 | SE-ResNet-152 | 0.787±0.007 | 0.900±0.004 | EfficientNet-B3 | 0.767±0.034 | 0.907±0.018 |
| VGG-19 | 0.782±0.016 | 0.882±0.031 | SENet-154 | 0.782±0.040 | 0.892±0.004 | EfficientNet-B4 | 0.779±0.030 | 0.905±0.022 |
| ResNet-18 | 0.774±0.037 | 0.900±0.007 | DenseNet-121 | 0.784±0.009 | 0.900±0.004 | EfficientNet-B5 | 0.769±0.018 | 0.905±0.020 |
| ResNet-34 | 0.790±0.025 | 0.910±0.018 | DenseNet-169 | **0.795±0.009** | **0.905±0.020** | EfficientNet-B6 | 0.767±0.037 | 0.890±0.010 |
| ResNet-50 | **0.792±0.014** | **0.910±0.016** | DenseNet-201 | 0.777±0.020 | 0.902±0.021 | EfficientNet-B7 | 0.794±0.033 | 0.877±0.009 |
| ResNet-101 | 0.779±0.025 | 0.905±0.030 | DenseNet-161 | 0.772±0.028 | 0.902±0.021 | | | |
| ResNet-152 | 0.754±0.052 | 0.905±0.012 | EfficientNet-B0 | 0.792±0.009 | 0.912±0.007 | | | |

Values are presented as Means ± S.D. Accuracy with DA: AD vs. CN classification using TRRA, Accuracy without DA: AD vs. CN classification without applying data augmentation.

## Appendix B. Classification performance of different data augmentation strategies

Table B.1 shows AD classification accuracy of EfficientNet-B1 with different data augmentation strategies on ADNI validation set. We can observe that accuracy first increases to the maximum value and then decreases as the value of $N$ increases for all data augmentation strategies. The decrease in accuracy is most likely because too many transformations are superimposed on the input image, which l leads to an inherent characteristics gap between the augmented image and the original image.

Table B.1 AD classification accuracy of EfficientNet-B1 with different data augmentation strategies on ADNI validation set.

| Method | $M$ | $N$ | | | | | | | |
|---|---|---|---|---|---|---|---|---|---|
| | | 1 | 2 | 3 | 4 | 5 | 6 | 7 | 8 |
| RA | 5 | 0.882±0.026 | 0.895±0.032 | 0.892±0.023 | 0.892±0.025 | 0.900±0.025 | 0.890±0.015 | **0.902±0.021** | 0.900±0.023 |
| | 10 | 0.867±0.032 | 0.895±0.027 | 0.892±0.019 | 0.887±0.027 | 0.897±0.030 | 0.897±0.026 | 0.900±0.023 | 0.900±0.023 |
| | 15 | 0.872±0.028 | 0.885±0.029 | 0.895±0.021 | 0.895±0.027 | 0.900±0.028 | 0.895±0.021 | 0.897±0.030 | 0.900±0.028 |
| | 20 | 0.875±0.026 | 0.892±0.023 | 0.892±0.023 | 0.892±0.023 | 0.897±0.030 | 0.895±0.027 | 0.895±0.022 | 0.900±0.028 |
| | 25 | 0.877±0.034 | 0.890±0.002 | 0.888±0.026 | 0.900±0.023 | 0.897±0.030 | **0.902±0.021** | 0.895±0.022 | 0.900±0.023 |
| | 30 | 0.867±0.034 | 0.885±0.029 | 0.892±0.023 | 0.892±0.023 | 0.895±0.032 | 0.890±0.025 | 0.890±0.020 | 0.892±0.023 |
| RA-23 | 5 | 0.880±0.032 | 0.882±0.026 | 0.887±0.021 | 0.890±0.025 | 0.897±0.025 | 0.902±0.021 | 0.895±0.021 | 0.902±0.021 |
| | 10 | 0.875±0.040 | 0.890±0.022 | 0.892±0.028 | 0.897±0.025 | 0.902±0.028 | 0.902±0.022 | **0.905±0.020** | 0.902±0.021 |
| | 15 | 0.884±0.032 | 0.880±0.022 | 0.882±0.013 | 0.887±0.021 | 0.900±0.018 | 0.902±0.022 | 0.902±0.021 | 0.902±0.022 |
| | 20 | 0.872±0.022 | 0.877±0.023 | 0.897±0.031 | 0.895±0.021 | 0.900±0.013 | **0.905±0.015** | 0.902±0.012 | 0.902±0.012 |
| | 25 | 0.872±0.032 | 0.882±0.025 | 0.892±0.023 | 0.900±0.018 | 0.902±0.012 | 0.902±0.016 | 0.895±0.006 | 0.900±0.015 |
| | 30 | 0.877±0.033 | 0.882±0.026 | 0.895±0.022 | 0.880±0.018 | 0.900±0.015 | 0.897±0.013 | 0.900±0.009 | 0.885±0.020 |
| RRA-23 | [5, 10] | 0.872±0.022 | 0.875±0.032 | 0.890±0.030 | 0.895±0.027 | 0.895±0.011 | 0.900±0.019 | 0.902±0.012 | 0.905±0.022 |

| | | | | | | | | |
|---|---|---|---|---|---|---|---|---|
| [5, 15] | 0.870±0.025 | 0.875±0.040 | 0.882±0.022 | 0.885±0.012 | 0.892±0.019 | 0.897±0.021 | 0.897±0.020 | 0.897±0.094 |
| [5, 20] | 0.872±0.021 | 0.889±0.022 | 0.895±0.032 | 0.887±0.021 | 0.897±0.020 | 0.892±0.015 | 0.902±0.016 | 0.900±0.094 |
| [5, 25] | 0.865±0.030 | 0.882±0.020 | 0.882±0.020 | 0.887±0.021 | 0.897±0.020 | 0.897±0.021 | 0.895±0.021 | 0.900±0.007 |
| [5, 30] | 0.872±0.021 | 0.880±0.016 | 0.880±0.021 | 0.887±0.021 | 0.897±0.020 | 0.895±0.012 | **0.907±0.013** | 0.900±0.023 |

Values are presented as Means ± S.D

## Appendix C. Classification performance of TRRA

Table C.1 shows AD classification accuracy of EfficientNet-B1 with TRRA under different hyperparameters on ADNI validation set. From the results we observe that randomly selecting five transformations from the $[color]$ category and randomly selecting two transformations from the $[shape]$ category achieve the best performance, which indicates that $[color]$ category contributes more to classification performance than $[shape]$ category.

Table C.1 AD classification accuracy on ADNI validation set of TRRA.

| $N_{color}$ | $N_{shape}$ | P | | | | | |
|---|---|---|---|---|---|---|---|
| | | 0.1 | 0.3 | 0.5 | 0.7 | 0.9 | 1 |
| 1 | 6 | 0.870±0.031 | 0.882±0.037 | 0.872±0.022 | 0.885±0.025 | 0.882±0.015 | 0.885±0.023 |
| 2 | 5 | 0.862±0.034 | 0.880±0.028 | 0.892±0.029 | 0.888±0.028 | 0.887±0.022 | 0.900±0.029 |
| 3 | 4 | 0.880±0.031 | 0.887±0.028 | 0.892±0.023 | 0.902±0.021 | 0.905±0.020 | 0.897±0.009 |
| 4 | 3 | 0.880±0.034 | 0.890±0.025 | 0.902±0.022 | 0.900±0.018 | 0.910±0.018 | 0.912±0.009 |
| 5 | 2 | 0.852±0.023 | 0.890±0.026 | 0.902±0.016 | 0.912±0.022 | **0.915±0.018** | **0.915±0.018** |
| 6 | 1 | 0.870±0.020 | 0.900±0.022 | 0.902±0.016 | 0.900±0.009 | 0.910±0.018 | 0.910±0.022 |

Values are presented as Means ± S.D.